\documentclass[prl,aps,english,twocolumn,amsmath,amssymb,superscriptaddress]{revtex4-2}
\usepackage{epsfig, graphicx,graphics,amsmath,amssymb,float}
\usepackage[T1]{fontenc}
\usepackage[latin9]{inputenc}
\usepackage{amsmath}
\usepackage{amssymb}

\usepackage{amscd}
\usepackage{bm}
\usepackage{psfrag}
\usepackage{bbm,dsfont} 
\usepackage{babel}
\usepackage{wasysym }
\usepackage{mathrsfs}
\usepackage{color}

\newcommand{\be}{\begin{equation}}
\newcommand{\ee}{\end{equation}}
\newcommand{\bea}{\begin{eqnarray}}
\newcommand{\eea}{\end{eqnarray}}
\usepackage{braket} 
\newcommand{\mc}{\mathcal}
\newcommand{\mb}{\mathbf}

\usepackage[final]{hyperref} 
\hypersetup{
	colorlinks=true,       
	linkcolor=blue,        
	citecolor=blue,        
	filecolor=magenta,     
	urlcolor=blue         
}

\newcommand{\new}[1]{\textcolor{black}{#1}}

\begin{document}

\title{Local spin-flip transitions induced by magnetic quantum impurities in two-dimensional magnets}

\author{Tim Bauer}
\affiliation{Institut f\"ur Theoretische Physik,
Heinrich-Heine-Universit\"at, D-40225  D\"usseldorf, Germany}

\author{Lucas R. D. Freitas} 
\affiliation{International Institute of Physics and Departamento de F\'isica Te\'orica
e Experimental, Universidade Federal do Rio Grande do Norte, 
Natal, RN, 59078-970, Brazil}

\author{Eric C. Andrade}
\affiliation{Instituto de F{\'i}sica, Universidade de S\~{a}o Paulo, 05315-970 S\~{a}o Paulo, SP, Brazil}

\author{Reinhold Egger}
\affiliation{Institut f\"ur Theoretische Physik,
Heinrich-Heine-Universit\"at, D-40225  D\"usseldorf, Germany}

\author{Rodrigo G. Pereira}
\affiliation{International Institute of Physics and Departamento de F\'isica Te\'orica
e Experimental, Universidade Federal do Rio Grande do Norte, 
Natal, RN, 59078-970, Brazil}

\begin{abstract}
We predict a general local spin-flip transition mechanism caused by magnetic quantum impurities 
in (partially) polarized phases of quantum magnets in the absence of conservation laws.  
This transition arises when a magnon bound state crosses zero energy as function of the magnetic field.   
As application, we study 2D van der Waals magnets described by the Kitaev-Heisenberg  honeycomb model which applies to the
transition metal trihalides CrI$_3$ and $\alpha$-RuCl$_3$.  We consider adatom and substitutional impurity positions, 
and show how spin-flip transitions can be detected in scanning tunneling spectroscopy.  
\end{abstract}
\maketitle

\emph{Introduction.---}Understanding the fascinating properties of the recently discovered 2D van der Waals magnets is a topic of enormous current interest; for reviews, see \cite{Burch2018,Gibertini2019,Khan2020,Yang2021rev,Kurebayashi2022,Wang2022rev,Ahn2024}.  These materials can be directly studied by surface probe techniques. In particular, their local magnetization profile has been mapped out by scanning nitrogen-vacancy magnetometry and by magnetic force microscopy \cite{Ahn2024}.  With the atomic resolution offered by
scanning tunneling spectroscopy (STS) \cite{Yin2021}, one may examine single-atom spin-flip processes  and spin excitations localized near impurities in full detail \cite{Heinrich2004,Wiesendanger2009}.
In this Letter, we study magnetic quantum impurities in 
(partially) polarized phases of 2D magnets in the absence of conservation laws. 
 We uncover a mechanism for local spin-flip transitions tied to sub-gap magnon bound states induced 
by magnetic quantum impurities.  Whenever the bound state energy crosses zero as function of the magnetic field, 
we predict such a transition to happen. This mechanism differs from a conventional spin-flop 
transition \cite{Blundell}.  We also develop a low-energy continuum approach showing that these spin-flip transitions 
appear not only in 2D magnets but also in 1D spin chains, and possibly even in 3D magnets.

For concrete calculations, we focus on the Kitaev-Heisenberg (KH) honeycomb model for transition metal trihalides \cite{Ahn2024}.  This material class includes CrI$_3$, a celebrated  2D ferromagnet with a small spin gap due to anisotropic exchange interactions \cite{Huang2017,Lado2017,Chen2020,Lee2020,Chen2021,Stavropoulos2021}. 
A second example is the layered Kitaev material $\alpha$-RuCl$_3$   \cite{Winter2017rev,Knolle2019,Hickey2021,Rousochatzakis2024}. In the latter, the effects of dilute Cr$^{3+}$ magnetic impurities  have been linked to the Kondo screening by a Majorana metal phase \cite{Lee2023}, but magnon bound states may also  potentially affect  the low-energy spectrum.
We note that STS observations of bound states induced by magnetic impurities in the spin-liquid candidate TaSe$_2$ have been interpreted  as spinon-Kondo effect \cite{Chen2022,He2022}. 
  Similar bound states have been studied in the context of the Kondo effect in spin chains \cite{Kattel2024}, a problem of relevance also for nanographene chains \cite{Mishra2021,Jacob2021}. 

For 2D magnets in a magnetic field, non-magnetic impurities (such as vacancies or bond defects) are generally \emph{not} expected to induce bound states below the magnon gap. For instance, for the 2D  KH model
\cite{Kitaev2006,Savary2017,Zhou2017,Hermanns2018,Takagi2019,Motome2020,Trebst2022} 
in a partially polarized phase \cite{McClartyPRB2018,Chern2021,Zhang2021b}, non-magnetic impurities generate magnon bound states only inside the energy gap between magnon bands of opposite Chern number \cite{Mitra2023,Bauer2023}. These bound states are precursors of the chiral edge states in topological magnon phases \cite{Slager2015,Diop2020,Chern2021,Zhang2021b,Habel2023}. On the other hand,
for a classical magnetic impurity (with spin $S_{\rm imp}\gg 1$), which is equivalent to a local magnetic field, bound states below the magnon gap have been  predicted \cite{Bauer2023}. This effect is independent of the band topology,
 see the Supplementary Material (SM) \cite{SM}.

We here include quantum fluctuations of the magnetic impurity by using exact diagonalization (ED) for small lattices, linear spin wave (LSW) theory in the thermodynamic limit, and a low-energy continuum theory.  Quantum effects are shown to \emph{qualitatively} change the scaling properties of the sub-gap bound state energy.  Depending, in particular, on the values of the bulk ($S$) and the impurity ($S_{\rm imp}$) spins and on the impurity position type (e.g., adatom vs substitutional), the bound state energy can reach zero multiple times as function of the magnetic field.  
Whenever this happens, we predict a \emph{discontinuity} in the local magnetization of the impurity and/or its neighboring bulk spins.  
Such local  spin-flip transitions manifest themselves as pronounced steps in the field dependence of the zero-bias STS conductance.  By scanning the STS conductance at finite bias voltage, one obtains information about the 
bound state energy.  Moreover, by performing spatial STS scans, the impurity type can be fully resolved.   

\begin{figure}
\begin{center}
   \includegraphics[width= \columnwidth]{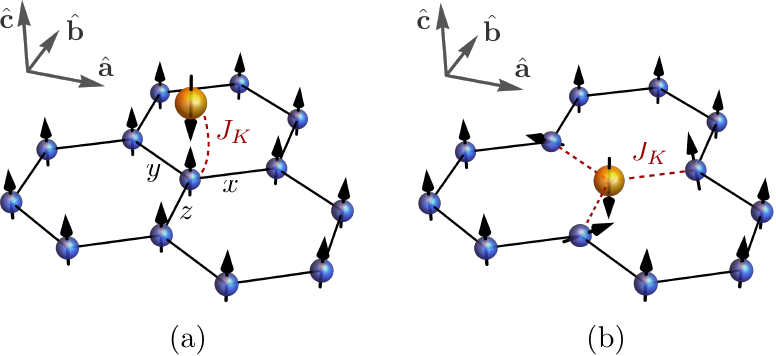}
    \caption{ Sketch of a (small part of a) 2D magnet with honeycomb lattice structure (bulk spins are shown in blue) and a single magnetic quantum impurity (yellow).
    Coordinate unit vectors $(\hat{\mb a},\hat {\mb b},\hat{\mb c})$ and nearest-neighbor bond types $(x,y,z)$ are also indicated. We assume $\mb h=h\hat{\mb c}.$
    (a) Adatom position of the impurity with isotropic exchange coupling $J_K>0$ to a single bulk spin.
    (b) Substitutional position, where the impurity replaces one bulk spin and interacts through isotropic couplings $J_K>0$ with the three neighboring bulk spins. Here we schematically illustrate that, in general, the classical spin configuration minimizing the energy can be inhomogeneous due to spin canting in the vicinity of the impurity \cite{SM,villain79,wollny11,wollny12,maryasin13}.
    }
    \label{fig1}
\end{center}
\end{figure}

\emph{General setup.---}We consider 2D spin Hamiltonians of the form ($\hbar=e=1$ below)
\be\label{model}
 H=\sum_{\langle j,k\rangle} \mb S^T_j H_{j,k} \mb S^{\phantom{T}}_k-\sum_j\mb h_{j}\cdot \mb S_j,
 \ee
 where $\mb S_j=(S_j^x,S_j^y,S_j^z)^T$ is a spin operator at site $j$, $\langle j,k\rangle$ denotes a bond between neighboring sites, $H_{j,k}$ are spin-spin coupling matrices, and  
 the vector $\mb h_j$ is proportional to the local magnetic field at site $j$.    
 For simplicity, we consider a homogeneous field $\mb h=h\hat{\mb c}$, see Fig.~\ref{fig1}.
 The bulk sites are occupied by spins $S$, and there is a single magnetic impurity with spin $S_{\rm imp}$.   We denote the impurity site by $j=0$, which is included in the summations in Eq.~\eqref{model}.  
 The bulk spins reside on a honeycomb lattice, where we consider two types of impurity locations as illustrated in Fig.~\ref{fig1}. 
 In the \emph{adatom} case, the magnetic impurity couples to a single bulk spin. 
 In the \emph{substitutional} case, one replaces a single bulk spin by the impurity spin which then couples to the three neighboring bulk spins.  For other configurations, e.g., if the impurity is located in the center of a hexagonal plaquette and thus couples to six bulk spins, results can be obtained by adapting our calculations for the cases considered here. 
For anisotropic exchange interactions, spin-rotational symmetry is absent, but the bulk spin coupling matrix $H_{j,k}$ can be constrained by the space group symmetry of the lattice \cite{Rau2018}. As paradigmatic model for transition metal trihalides, we consider the KH honeycomb model \cite{Janssen2019} for bulk spin $S=3/2$ and ferromagnetic bulk exchange coupling $J_b>0$.  For CrI$_3$, first-principles calculations and analysis of experimental data \cite{Xu2018,Ubiergo2021,Cen2023} indicate a subdominant antiferromagnetic Kitaev coupling $K<0$. 
In the SM, we also present results for the idealized ferromagnetic ($K>0$, $J_b=0$) Kitaev model with $S=1/2$ \cite{Kitaev2006}. We note that it is straightforward to generalize our 
approach to include other couplings, e.g., Dzyaloshinskii-Moriya interactions between next-nearest neighbors. 

For the impurity-bulk spin couplings in Fig.~\ref{fig1}, we instead consider an isotropic antiferromagnetic exchange coupling,  $H_{0,j} =J_K\mathbbm{1}_3$ with $J_K>0$. This is a natural assumption if the impurity atom has no orbital degeneracy \cite{Lee2023}. 
Next, we recall that for a free magnetic ion with spin $S$, orbital angular momentum $L$, and total 
angular momentum $J$, the Land\'e factor is $g_L=\frac32+\frac{S(S+1)-L(L+1)}{2J(J+1)}$ \cite{FazekasBook}. This value is typically a good approximation for $4f$ ions like Yb$^{3+}$, where crystal field effects are weak and $J$ follows from Hund's rules.
On the other hand, for $3d$ ions like Cr$^{3+}$, $L$ is quenched by crystal field effects and hence $g_L=2$ for all  $S$.  
For Ru$^{3+}$ ions in $\alpha$-RuCl$_3$, the intricate 
interplay between orbital degeneracy, crystal field, and spin-orbit coupling \cite{Jackeli2009} implies that $g_L$ is an anisotropic tensor \cite{FazekasBook}, \new{with $g_L\approx 1.3$ for magnetic fields along the $\hat{\mb c}$ direction} \cite{Winter2018}. 
Assuming a homogeneous external magnetic field, we absorb 
the Bohr magneton and the bulk Land{\'e} factor into $\mb h_j=\mb h$ in Eq.~\eqref{model} for bulk sites $(j\ne 0$). 
For the impurity spin ($j=0$), we set 
\begin{equation}\label{relative}
    \mb h_0=g\mb h,\quad g=g_L^{\rm imp}/g_L^{\rm bulk},
\end{equation}
where $g$ is the relative Land{\'e} factor of the impurity compared to the bulk spins.  
For instance, for Ru$^{3+}$ ions and ${\mb h}=h\hat{\mb c}$, one finds $g\approx 1.5$ and $S_{\rm imp}=1/2$ both for Co adatoms \cite{Chen2022} and for Ti$^{3+}$ ions at substitutional sites. 

Let us first summarize several key aspects.
We assume that, without the impurity, the system is in a gapped (partially) polarized phase with magnons as low-energy excitations, 
where the magnon gap is basically set by the magnetic field, and study sub-gap magnon bound states induced by a single magnetic quantum impurity. 
The impurity coupling $J_K$ now competes with the magnetic field $\mb h$. 
In a classical picture, $\mb h$ tries to polarize all spins in the same direction, but spins 
coupled by $J_K$ align in opposite directions if $J_K\gg |\mb h|$.  For the adatom case with $S_{\rm imp}=S$, 
the limit $J_K\to \infty$  corresponds to singlet formation between the impurity and a bulk spin, where both spins are frozen out and can be described as a vacancy.  However, for the KH honeycomb model, a vacancy does \emph{not} induce sub-gap bound states \cite{Mitra2023,Bauer2023}. As one varies $J_K/|\mb h|$ between the weak- and strong-coupling limits, either the impurity or the bulk spin has to flip against the magnetic field. In the adatom case with $S_{\rm imp}\neq S$ and/or in the substitutional case, the strong-coupling limit may have a residual spin, and thus multiple transitions are possible. 
Below we describe such discontinuous spin transitions in the quantum case and show how they can be detected in STS.
  
\begin{figure}
\begin{center}
\includegraphics[width= 0.98\columnwidth]{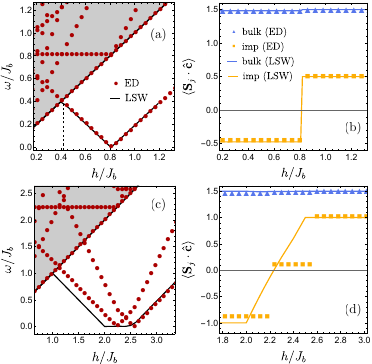}
\caption{ED and LSW results for the KH model with $S=3/2$, $K/J_b=-0.2$, 
$J_K/J_b=0.5$, and $\mb h=h\hat{\mb c}$. For all cases studied here, $g=1$ in Eq.~\eqref{relative}. The shaded region indicates the magnon continuum $\omega>\omega_g$ in the thermodynamic limit, with the magnon gap $\omega_g=h$.  ED results refer to a cluster with $2\times 2$ unit cells (9 sites for the adatom case) and periodic boundary conditions. 
\emph{Adatom impurity:} (a) Excitation spectrum $\omega/J_b$ (relative to the ground state) vs magnetic field $h/J_b$ for $S_{\rm imp}=1/2$.  ED (LSW) results correspond to red dots (black lines). The dashed vertical line indicates the classical spin flip transition.
(b) Spin projections $\langle{\mb S}_{0,1}\cdot \hat{\mb c}\rangle$ vs $h/J_b$ for the case in panel (a).
Yellow squares (lines) correspond to ED (LSW) results for the impurity spin. Results for the
coupled bulk spin are shown as blue triangles (ED) and blue lines (LSW), respectively.
\emph{Substitutional impurity:} (c) $\omega/J_b$ vs $h/J_b$ and (d) $\langle{\mb S}_{0,1}\cdot \hat{\mb c}\rangle$ vs $h/J_b$ for $S_{\rm imp}=1$. 
\label{fig2}}
\end{center}
\end{figure}

\emph{KH model.---}The bulk spin couplings $H_{j,k}$ in Eq.~\eqref{model} include an isotropic ferromagnetic Heisenberg coupling $J_b>0$, and a bond-dependent Kitaev contribution $K$. For instance, for a $z$-bond in Fig.~\ref{fig1}, we have $H_{j,k}=-{\rm diag}(J_b,J_b,J_b+K)$.  
Bulk interactions for bonds of type $\gamma\in\{x,y\}$ then follow by cyclic permutation of $(\alpha,\beta,\gamma)$ in $H^{\alpha,\beta}_{j,k}$.
In CrI$_3$,  bulk spins (Cr$^{3+}$ ions) have $S=3/2$ \cite{Wang2022rev}. 
Magnetic impurities at substitutional sites could be, e.g., V$^{3+}$ ions with $S_{\rm imp}=1$ \cite{Son2019}
where we obtain $g=1$ from Eq.~\eqref{relative}. 
On the other hand, adatom impurities should be neutral, e.g., 
Co atoms with $S_{\rm imp}=1/2$ \cite{Chen2022},  again resulting in $g=1$.
Ab-initio calculations for CrI$_3$ \cite{Yang2021a} favor impurity locations in the hexagon center,  
while for Ni atoms, the adatom location is also possible.  
However, with a scanning probe tip, one can move around impurities at will \cite{Chen2022}.  

We show ED and LSW results for $S=3/2$ in Fig.~\ref{fig2}.  
While ED is a numerically exact method, it is limited to small system size.  
 LSW theory instead allows us to treat the thermodynamic limit, where the first step is to determine the classical spin configuration that minimizes the energy.  In general, there can be spin canting as schematically illustrated in Fig.~\ref{fig1}(b).  One then performs a Holstein-Primakoff 
transformation to boson operators describing the magnons. In principle, the linearized theory
becomes exact for large $S$ \cite{FazekasBook,Chern2021,Zhang2021b,Bauer2023}.
Since the impurity breaks translation invariance, one diagonalizes the quadratic LSW Hamiltonian in real space.  This can be done for fairly large lattices, allowing for an extrapolation to the thermodynamic limit \cite{SM}. 

Remarkably, both methods give almost perfect agreement for an  adatom impurity with $S_{\rm imp}=1/2$, see Fig.~\ref{fig2}(a,b). Here a sub-gap magnon bound state appears whose energy vanishes for $h/J_b\approx 0.8$. 
For $h\ll J_K$, the bound state merges with the magnon continuum as expected from the impurity screening scenario discussed above. 
As seen in Fig.~\ref{fig2}(b), the spin projections $\langle \mb S_{0,1}\cdot \hat {\mb c}\rangle$ of the impurity spin and (to much lesser degree) the coupled bulk spin exhibit \emph{discontinuous} jumps at $h/J_b\approx 0.8$.  
Importantly, the transition obtained within LSW theory is due to quantum fluctuations on top of a uniformly polarized spin configuration. By contrast, a classical transition might be expected when the energy of the configuration with the flipped impurity spin falls below the energy of the uniform configuration. Calculating these energies within LSW theory up to order $S$ (in the $1/S$ expansion) \cite{SM}, we find that this semiclassical criterion underestimates the critical magnetic field, see the dashed vertical line in Fig.~\ref{fig2}(a).

We next address the \emph{substitutional} case for $S_{\rm imp}=1$, see Fig.~\ref{fig2}(c,d),
where marked differences between ED and LSW results emerge.  In particular, ED results indicate \emph{two} spin transitions associated with multiple zeros of the bound state energy. On the other hand, the LSW spectrum calculated for the uniform classical spin configuration (with polarized impurity) captures only one transition. 
In Fig.~\ref{fig2}(c,d), we show the LSW spectrum and spin projections for a configuration in which the impurity spin is allowed to rotate to minimize the classical energy  \cite{SM}. This continuous rotation  of the impurity spin breaks the exact C$_3$ lattice symmetry.
However, such a spontaneous symmetry breaking is not possible for a single degree of freedom with local interactions, since there are only finite energy barriers to other broken-symmetry states and quantum fluctuations can restore the symmetry. In fact, we never observed transverse components of the impurity spin in ED. Another indication of the failure of LSW theory is that the predicted magnon bound state energy vanishes in the regime where the classical spin rotates (which is where ED identifies multiple transitions). In this case, the putative classical configuration cannot be stable because one can add magnons without energy cost. Given the excellent agreement between ED and LSW results in Fig.~\ref{fig2}(a,b), we believe that ED captures the thermodynamic limit also in Fig.~\ref{fig2}(c,d). 
The transitions at $h/J_b\approx 2.2$ and $h/J_b\approx 2.55$ are again predominantly due to discontinuities in the local magnetization. They correspond to zeros of the energy for excitations with $n$ magnons in the bound state, where $n\leq 2S_{\rm imp}$. For higher $S_{\rm imp}$, one can thus expect more spin transitions.

\emph{Low-energy continuum approach.---}For the adatom case and ${\mb h}=h\hat{\mb c}$, we can analytically determine the scaling of the sub-gap magnon bound state energy $E_b<0$ (relative to the magnon gap $\omega_g$) in the regime $J_K\ll h$, where Eq.~\eqref{model} affords an effective continuum description 
\cite{SM}. The bulk magnon dispersion can be approximated by $\omega(\mb k)\approx \omega_g+\frac{\mb k^2}{2m}$.
(For the KH model, $\omega_g=h$ and $m^{-1}=J+\frac{K}{3}-\frac{K^2}{h+3J+K}.$)
The low-energy Schr\"odinger equation for a single magnon described by $\psi(\mb r)$ with $\mb r=(r_x,r_y)^T$ then contains 
an attractive $\delta$-function potential due to the impurity, 
\be\label{schr2}
-\frac{1}{2m}\nabla^2\psi(\mb r) -V_{\rm eff}(E_b) \delta(\mb r)\psi(\mb r) =E_b\psi(\mb r), 
\ee
with the energy-dependent coupling strength
$V_{\rm eff}(E_b)=J_K S_{\rm imp} \left(1 +\frac{J_K S  }{\varepsilon_a+|E_b|}\right),$
where $\varepsilon_a$ is an energy scale for the impurity.
(For the KH model, $\varepsilon_a=gh-\omega_g - J_K S$.)
For a classical magnetic impurity with $S_{\rm imp}\to \infty$ at constant $J_K S_{\rm imp}$ \cite{Shiba1968,Imry1975,Bauer2023}, we obtain the nonperturbative scaling law $E_b\propto - e^{-c_b/(mJ_K)}$ \cite{Jackiw1991}, where $c_b$ is of order unity.  However, for a quantum impurity, we find a qualitatively different scaling, $E_b\simeq gh-\omega_g-J_KS+{\cal O}(J_K^2)$, in accordance with our numerical ED and LSW results.  By varying the magnetic field, the bound state crosses zero energy (i.e., $E_b=-\omega_g$), and a discontinuous local spin-flip transition occurs.   The low-energy equation \eqref{schr2} is not limited to the KH model but applies to many other 2D magnets as well.  With minor modifications, it also 
describes 1D and 3D magnets, underlining the generality of the found spin-flip mechanism. 
This mechanism is qualitatively different from a conventional 
spin flop transition \cite{Blundell} which involves a first-order transition without  magnon gap closing, see Ref.~\cite{SM}. 

\begin{figure}
\begin{center}
\includegraphics[width= 0.98\columnwidth]{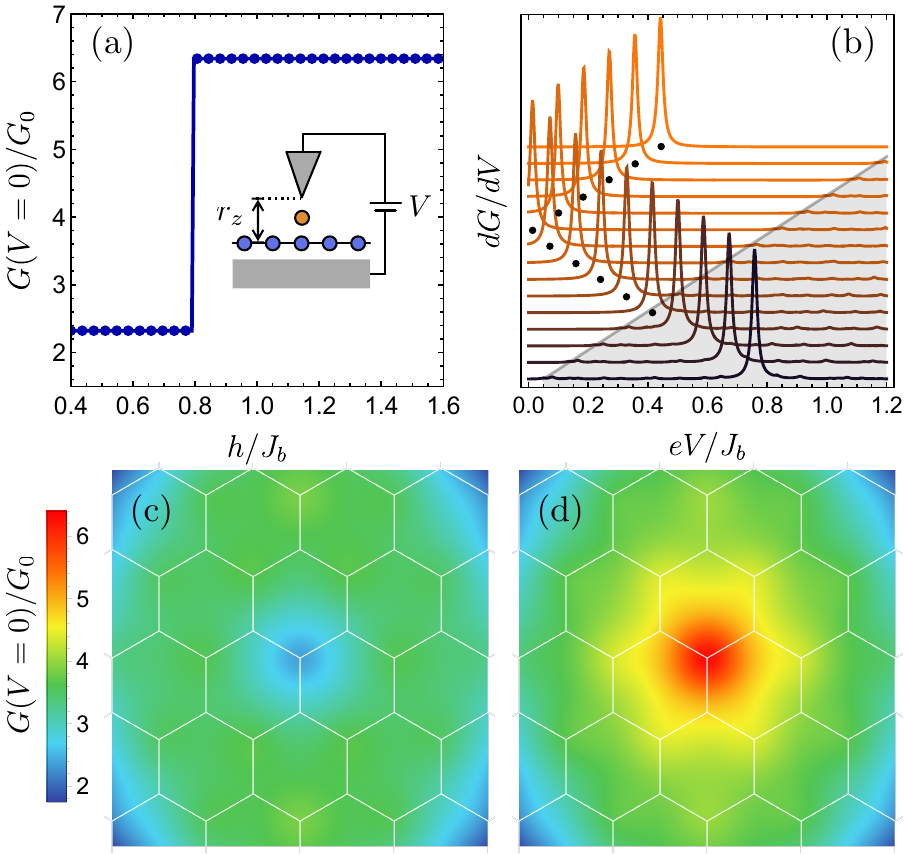}
    \caption{STS conductance for the KH model with an adatom magnetic impurity from LSW theory, cf.~Fig.~\ref{fig2}(a,b), with $K=-0.2J_b$, $J_K=0.5J_b$, $S=3/2$, $S_{\rm{imp}}=1/2$, and $g=1$. The reference conductance
    $G_0=2\pi d_Ad_B |t_{\rm bulk}|^2$ involves densities of states $d_A$ and $d_B$ for tip and substrate, 
    and we use cotunneling amplitudes $t_{\rm imp}$ and $t_{\rm bulk}$ for the impurity and bulk spins, respectively. 
    The STM tip and the impurity position are characterized by $l_0=r_z=0.5a_0$, $t_{\rm{imp}}/t_{\rm{bulk}}=1.5$, and $\mb{R}_0=\mb{R}_1+0.2a_0\hat{\mb c}$, where $a_0$ is the lattice constant and $l_0$ the tip resolution.
    The bulk spin at site $\mb R_1$ is coupled to the impurity at site $\mb R_0$. (a) $V=0$ conductance vs magnetic field $h$ for a tip above the impurity as schematically depicted in the inset. (b) Conductance derivative $dG/dV$ (in arbitrary units) vs bias voltage. Different curves correspond to uniformly spaced field values ranging from $h=0.05J_b$ (bottom) to $h=1.25J_b$ (top curve), shifted vertically to aid visualization.  The $\delta$-function peaks have been broadened by a Lorentzian of width $0.01J_b$. The dots track the bound state energy.  (c,d) Color-scale plots for the spatial profile of the zero-bias conductance near the magnetic impurity (which is in the center of the respective panel) for two field values: (c) $h=0.65J_b$ (below the transition), (d) $h=1.164J_b$ (above the transition). }
    \label{fig3}
\end{center}
\end{figure}

\emph{STS conductance.---}We next discuss how the above spin transitions can be observed in STS. Known expressions for the zero-temperature STS conductance $G(V,\mb r_t)$  at bias voltage $V$ and tip position ${\mb r}_t=(r_x,r_y,r_z)^T$ ~\cite{fransson2010theory,fernandez2009theory,Feldmeier2020,Carrega2020,Mitra2023,Bauer2023,Kao2024a,Kao2024b} are summarized in the SM \cite{SM}, involving a form factor and the 
dynamical spin correlation function of the 2D magnet. (Here the $z$-axis is aligned with  $\hat{\mb{c}}$.) The zero-bias STS conductance is expressed by the spin expectation values $\langle S_j^\alpha \rangle$, and can thus directly reveal spin transitions.
We illustrate the STS conductance in Fig.~\ref{fig3} for an adatom impurity with $S_{\rm imp}=1/2$, cf.~Fig.~\ref{fig2}(a,b).  The step in the field dependence of the zero-bias conductance seen in Fig.~\ref{fig3}(a) directly follows from the spin transition in Fig.~\ref{fig2}(b). 
At low but finite bias voltage, Fig.~\ref{fig3}(b) illustrates that the conductance derivative $dG/dV$ exhibits a pronounced \emph{peak} whenever the voltage matches the sub-gap magnon bound state energy.  By varying both the magnetic field and the bias voltage, one can thus map out the field dependence of the bound state energy.  
Finally, \new{see} Fig.~\ref{fig3}(c,d), in spatial scans of the zero-bias conductance, the conductance switches from a minimum to a maximum near the impurity site when increasing $h/J_b$ across the transition at $h/J_b\approx 0.8$, \new{consistent with} Fig.~\ref{fig3}(a). 

\emph{Conclusions.---}Magnetic impurities in 2D magnets can give rise to magnon bound states at very low energies.  Such states \new{can cause similar behavior in STS as fractional excitations such as spinons or $\mathbb Z_2$ vortices in spin liquids. In order to distinguish such phases, one could characterize magnon bound states by STS by first exposing the system to a strong field. In a second step, one lowers the field to reach the putative spin liquid phase, where detailed STS predictions are available}  \cite{Bauer2023,Kao2024a,Kao2024b}. In any case, we expect that the predicted 
 spin transitions due to zero-energy magnon bound states will soon be observed.  

\begin{acknowledgments} 
We acknowledge funding by the Deutsche Forschungsgemeinschaft (DFG, German Research Foundation),
Projektnummer 277101999 - TRR 183 (project C01) and under Germany's Excellence Strategy - Cluster of Excellence Matter 
and Light for Quantum Computing (ML4Q) EXC 2004/1 - 390534769, by  the Simons
Foundation (Grant No. 1023171, R.G.P.), by the Brazilian ministries MEC and MCTI, by the Brazilian agencies CNPq and FAPESP, and by the Coordena\c{c}\~{a}o de Aperfei\c{c}oamento de Pessoal de N{\'i}vel Superior - Brasil (CAPES) - Finance Code 001.
\end{acknowledgments}

\bibliography{biblio}

\appendix
 
\section{Supplemental Material}

We here provide (i) details about linear spin wave theory in the presence of a single magnetic
quantum impurity, (ii) derive analytical results for the sub-gap magnon bound state energy in the weak-coupling limit, (iii) discuss the tunneling conductance expression, and (iv) show data for the idealized Kitaev honeycomb model. 
Equation (X) in the main text is referred to as Eq.~(MX) below.

Below we provide additional details and derivations concerning the results presented in the main text.  For concrete
results, we focus on a paradigmatic model for 2D van der Waals magnets like CrI$_3$, namely the 
Kitaev-Heisenberg \new{(KH)} model on the honeycomb lattice \cite{Janssen2019,Stavropoulos2021,Wang2022rev,Ahn2024}.
Leaving aside the so-called $\Gamma$ exchange terms \cite{Winter2017rev}, this model is also expected to 
describe the physics of $\alpha$-RuCl$_3$ layers. We discuss in Sec. I our formulation of linear spin wave (LSW) theory in the presence of a magnetic quantum
impurity for the Hamiltonian in Eq.~(M1).  As discussed in the main text, the corresponding results show excellent  agreement with exact diagonalization (ED) results for small systems in the adatom case with $S_{\rm imp}=1/2$.
For the substitutional impurity with $S_{\rm imp}=1$, however, we saw in the main text that the continuous   rotation of the impurity spin in the classical background of the LSW theory artificially breaks exact symmetries and can lead to spurious results. 
In Sec. II, we \new{derive} analytical results for the sub-gap magnon bound state energy which are valid in certain limits \new{and shown in the main text}.  Next, in Sec. III, \new{known expressions for the} scanning tunneling spectroscopy (STS) conductance are summarized, \new{including a brief outline of their derivation}.  
Finally, we address the idealized Kitaev honeycomb model in Sec. IV.

\section{I. LSW theory}\label{sec1}

As one method for studying partially polarized phases of 2D magnets with a single magnetic quantum impurity, see Eq.~(M1), we use LSW theory \cite{FazekasBook}.  One starts by determining the classical spin configuration 
minimizing the energy, see Sec. I\,A.  After rotating the local spin quantization axis to the corresponding classical spin axis, one performs a Holstein-Primakoff transformation to introduce boson operators for the magnon degrees of freedom.
We consider a linear approximation, where the Hamiltonian can be diagonalized by a Bogoliubov transformation, see Sec. I\,B. In Sec. I\,C, we briefly describe how one can obtain dynamical spin correlations from this approach, which 
enter the STS conductance \new{in Sec. III}.  In Sec. I\,D, we discuss estimates for the regime where discontinuous spin transitions occur based on comparing the energies of the competing classical configurations.  
\new{Finally, in Sec. I\,E, we contrast the found spin-flip transition mechanism from  
conventional spin-flop transitions \cite{Blundell}.}

\subsection{A. Classical spin configuration and spin canting}\label{sec1a}

\begin{figure}[t]
\begin{center}
\includegraphics[width=\columnwidth]{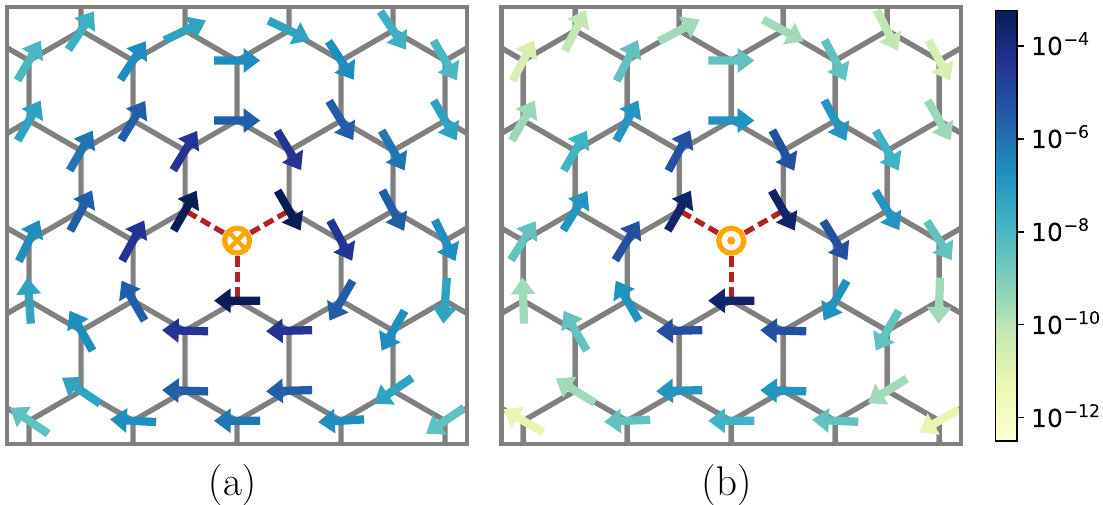} 
\end{center}
\caption{Examples for classical spin configurations near a magnetic impurity of substitution type in the KH model. Arrows represent spin components in the $ab$ plane, where the color scale indicates the magnitude of the in-plane component. Orange-colored symbols show the out-of-plane components of the impurity spin. Parameters are chosen as in Fig.~2(c) of the main text, i.e., $S=3/2$, $S_{\rm{imp}}=1/2$, $K/J_b=-0.2$, and $J_K/J_b=0.5$. Panel (a) shows results for $h/J_b=1.2$, i.e., below the spin transitions. 
Panel (b) is for $h/J_b=3.5$, i.e., above the spin transitions.  }
\label{figSM1}
\end{figure}

The first step is to find the classical spin configuration that minimizes the energy.  
In frustrated magnets, inhomogeneities can locally tip the energy balance and disturb the local spin environment, a phenomenon dubbed   local relief of frustration \cite{villain79}. Because a classical adatom magnetic impurity effectively acts as a local field, a local distortion (``spin canting'', schematically illustrated in Fig.~1(b) of the main text)  occurs in the neighboring bulk spins with respect to the uniform case.  For a substitutional impurity, the situation is similar, as the bulk spins neighboring the impurity feel an imbalance in their local exchange field.
In noncollinear long-ranged ordered phases, such local distortions decay as a power law in space and lead to nontrivial spin textures \cite{wollny11, wollny12, maryasin13}. In polarized phases, magnons are gapped and we expect an exponential decay of the texture, limiting it to the immediate vicinity of the impurity. 
We numerically determine the classical ground state by sequentially anti-aligning the spins with their local exchange field. To ensure convergence, we start from distinct initial conditions and select as the ground state the spin texture with the lowest energy. This procedure allows us to obtain canted spin texture and gauge the competition between the impurity-bulk exchange coupling $J_K>0$ and the external magnetic field strength $h$ in determining the local spin configuration.

As an example, we show results for the KH honeycomb model with a substitutional magnetic impurity. Figure \ref{figSM1} illustrates  the  classical spin configuration for the same parameters as in Fig.~2(c) of the main text. Here we choose two values of the magnetic field along the $\hat{\mb c}$ direction, where the impurity spin in the classical configuration is either antiparallel [Fig.~\ref{figSM1}(a)] or parallel [Fig.~\ref{figSM1}(b)] to the field direction. This way, we avoid the regime near the transitions where the classical state incorrectly predicts a smooth rotation of the impurity spin.   As shown in Fig.~\ref{figSM1},  we observe a vortex-like pattern for the bulk spin components in the $ab$ plane, both above and below the transition. In contrast to the spurious rotation of the impurity spin discussed in the main text, the vortex-like pattern for the bulk spins respects the 
discrete $\mathbb Z_3$ spin-rotation symmetry around the impurity site.  A similar pattern has been observed near a vacancy in the KH model \cite{Mitra2023}. 

\subsection{B. Holstein-Primakoff transformation}\label{sec1b}

Given the classical spin configuration, we apply a local rotation that aligns the $z$-axis with the local spin polarization, $\mb S_j=R_j\tilde{\mb S}_j,$ where $R_j$ is an orthogonal matrix. For polarized phases with a homogeneous spin configuration, $R_j=R$ is site independent and there is no spin canting. 
For instance, assuming that all spins are polarized by a magnetic field perpendicular to the honeycomb plane, $\mb h=h\hat{\mb c}$,  the spin basis used for representing the KH honeycomb model is rotated according to \cite{Kitaev2006}
\be
R = \begin{pmatrix}
    \frac{1}{\sqrt{6}}&\frac{-1}{\sqrt{2}}&\frac{1}{\sqrt{3}}\\
    \frac{1}{\sqrt{6}}&\frac{1}{\sqrt{2}}&\frac{1}{\sqrt{3}}\\
    \frac{-2}{\sqrt{6}}&\frac{1}{\sqrt{2}}&\frac{1}{\sqrt{3}}
\end{pmatrix}.
\ee
In any case, given the orthogonal matrices $R_j$, we next employ a standard Holstein-Primakoff transformation in order to represent spin operators in terms of bosonic operators $b_j$ and $b_j^\dagger$ describing magnons \cite{FazekasBook},
\bea\nonumber
\tilde S_j^z&=&S_j-n_j, \qquad  \tilde S_j^+= \sqrt{2S_j-n_j} \,b_j\approx \sqrt{2S_j} \,b_j, \\
\label{spins}\tilde S_j^-&=&b^\dagger_j \sqrt{2S_j-n_j} \approx \sqrt{2S_j}\,b_j^\dagger, 
\eea
where $n_j=b^\dagger_jb^{\phantom\dagger}_j$. The polarized impurity spin at $j=0$ is included in Eq.~\eqref{spins}, where $S_0=S_{\rm{imp}}$ generally differs from the value for bulk spins, $S_{j\neq 0}=S$.

To leading order in $1/S$, the Hamiltonian can be written as
\be \label{fullH}
H= E_{\rm cl}+H_{\rm sw}+{\mc O}\left(S^{1/2}\right),
\ee
where $E_{\rm cl}\sim {\mc O}\left(S^2\right)$ is the classical ground state energy and 
$H_{\rm sw}\sim {\mc O}\left(S\right)$ is the LSW Hamiltonian,
\be\label{lsw}
H_{\rm sw} = \sum_{\langle j,k\rangle }\left[t_{j,k}b^\dagger_j b^{\phantom\dagger}_k+\Delta_{j,k}b_jb_k+\text{h.c.} \right]
+ \sum_j h^{\rm{eff}}_j b^{\dagger}_jb^{\phantom\dagger}_j. 
\ee
For $S\to \infty$, the LSW approach, neglecting all terms beyond $H_{\rm sw}$ in Eq.~\eqref{fullH}, becomes formally exact (but see below and the main text for subtleties related to the classical reference configuration).
The parameters $t_{j,k}$ and $\Delta_{j,k}$ in Eq.~\eqref{lsw} follow from the rotated spin-spin interaction matrices, 
$\tilde{H}_{j,k}=R^T_j{H}_{j,k} R_k$ with $H_{j,k}$ in Eq.~(M1), as
\bea \nonumber
t_{j,k}&=&\frac{\sqrt{S_jS_k}}{2}\left[\left(\tilde{H}^{xx}_{j,k}-i\tilde{H}^{xy}_{j,k}\right)+\left(\tilde{H}^{yy}_{j,k}+i\tilde{H}^{yx}_{j,k}\right)
\right],\\
\Delta_{j,k} &=& \frac{\sqrt{S_jS_k}}{2}\left[\left(\tilde{H}^{xx}_{j,k}-i\tilde{H}^{xy}_{j,k}\right)-\left(\tilde{H}^{yy}_{j,k}+i\tilde{H}^{yx}_{j,k}\right)\right]\label{djk}
\eea
Note that the anomalous couplings $\Delta_{j,k}$ change the total number of magnons.
The effective field $h^{\rm{eff}}_j$ involves both the external field and a contribution from spin interactions. 
For bulk sites ($j\neq 0$), we find
\be\label{heffj}
    h^{\rm{eff}}_j=\tilde{h}^z-\sum_{k\in{\mc V}_j}S_k\tilde{H}^{zz}_{j,k},
\ee
where ${\mc V}_j$ denotes the set of nearest neighbors of $j$ and $\tilde{h}^z$ is the $z$-component of the rotated field $\tilde{\mb h}=R_j^T{\mb h}$.
At the impurity site $(j=0$), the effective field is given by
\be\label{heff0}
    h^{\rm{eff}}_0=g\tilde{h}^z-\sum_{k\in{\mc V}_0} S\tilde{H}^{zz}_{j,k}=gh-|{\mc V}_0|SJ_K,
\ee
where $|{\mc V}_0|=1$ for an adatom and $|{\mc V}_0|=3$ for a substitutional impurity. 
The second equality only holds for a polarized uniform spin configuration, where $\tilde{h}^z=h$.
The factor $g$ has been defined in Eq.~(M2).

The quadratic Hamiltonian \eqref{lsw} can be diagonalized by a Bogoliubov transformation. For a decoupled adatom impurity ($J_K=0$) and a fully polarized system, we can diagonalize the Hamiltonian in momentum space and obtain analytical expressions for the magnon dispersion relation. The magnon gap is given by $\omega_g=h$ for the KH model. For $J_K>0$, we perform a Bogoliubov transformation in real space for a finite system with periodic boundary conditions, and then extrapolate to the thermodynamic limit.  

As shown in Fig.~2 of the main text, sub-gap magnon bound states appear for arbitrarily small exchange coupling $J_K>0$. 
Within LSW theory, the discontinuous spin transition   is well described  by calculating $\langle \tilde S_j^z\rangle=S_j-\langle b^\dagger_j b^{\phantom\dagger}_j\rangle$ for the adatom case in the uniform classical configuration.   Even without changing the classical state, the discontinuity in the quantum correction to the magnetization is possible because there is a rearrangement of the eigenvectors in the Bogoliubov transformation when the lowest eigenvalue (associated with the magnon bound state) goes through zero. 

\subsection{C. Dynamical spin correlations}\label{sec1c}

Let us briefly sketch how to obtain dynamical spin correlations from this approach.
The Bogoliubov transformation yields the single-particle eigenenergies $\omega_n$ with 
bosonic eigenoperators $\tilde{b}_n$ and $\tilde{b}_n^\dagger$. We can thereby rewrite Eq.~\eqref{lsw} as
\be
\label{diagonalsw}
    H_{\rm sw}=\sum_n \omega_n\tilde{b}^\dagger_n\tilde{b}^{\phantom\dagger}_n+  \Delta E,\quad \Delta E=\frac12\sum_n\omega_n.
\ee
The Lehmann representation of the dynamical spin correlations appearing in the STS conductance expression, \new{see Eq.~\eqref{STScond} below,} is given by
\be
    C^{\alpha\beta}_{jk}(\omega)=\sum_\nu \braket{\Phi_0|S^\alpha_j|\Phi_\nu}\braket{\Phi_\nu|S^\beta_k|\Phi_0}\delta(\omega+E_0-E_\nu),
\ee
where $\ket{\Phi_\nu}$ are many-body eigenstates with  energy $E_\nu$, including the ground state $|\Phi_0\rangle$ with energy $E_0$.
Within LSW theory, the matrix elements $\langle\Phi_0|S^\alpha_j|\Phi_\nu\rangle$ are finite only for states of the form
\be
    \ket{\Phi_\nu}\propto\ket{\Phi_0},\,\ket{\Phi_\nu}\propto \tilde{b}^\dagger_n\ket{\Phi_0}\,\text{or }\ket{\Phi_\nu}\propto \tilde{b}^\dagger_n\tilde{b}^\dagger_m\ket{\Phi_0}.
\ee
We can thereby rationalize the emergence of zero-, one- and two-magnon contributions in the nonlinear STS conductance.
However,  we find that the two-magnon contribution of the bound state vanishes within LSW theory.  
For the pure Heisenberg model ($K=0$) in a perpendicular magnetic field above the transition, i.e., for $h\gg J_K$, this follows analytically from magnon number conservation, i.e., $\Delta_{j,k}=0$ in Eq.~\eqref{djk}. This fact implies that spin expectation values do not fluctuate for magnetic fields above the spin flip transition within LSW theory. This analytical argument does not hold below the transition or for finite Kitaev interaction, but we here have verified the absence of two-magnon contributions numerically.   

\subsection{D. Classical spin flip}\label{sec1d}

In the main text, we mentioned that LSW theory provides an estimate for a classical spin flip transition. At strong magnetic fields, $h\gg J_K$, the classical ground state  is  homogeneous and fully polarized. On the other hand, as we decrease the ratio $h/J_K$, at some point a spin flip will be energetically preferred in the classical ground state configuration. The energy cost of flipping the spin at site $j$ is given by $2S_j h^{\rm{eff}}_j$, with the effective field in Eq.~\eqref{heffj} for a bulk  spin and in Eq.~\eqref{heff0} for the impurity.  For an adatom impurity, the classical configuration with a flipped spin  has lower energy than the uniform classical state if the magnetic field falls below the critical value $h^*={\rm max}\left(h^*_{\rm{imp}},h^*_{\rm{bulk}}\right)$, where
\be\label{hstar}
     h^*_{\rm{imp}}=\frac{S}{g}J_K,\qquad h^*_{\rm{bulk}}=\sum_{k\in\mc V_{j'}}S_k\tilde{H}^{zz}_{j',k}
\ee
correspond to a spin flip of the impurity and the bulk spin $j'\in\mc V_0$ coupled to the impurity, respectively. Note that for $g\gg 1$, we have $h^*=h^*_{\rm{bulk}}$, meaning that in this case we expect the bulk spin to flip against the magnetic field as we decrease $h/J_K$.  

Applying Eq.~(\ref{hstar}) for an adatom impurity with $S_{\rm imp}=1/2$ and the same parameters as in Fig.~2(a,b) of the main text, we obtain the estimate $h^*/J_b\simeq 0.75$ for the classical spin flip transition. However, this estimate is based on comparing the classical energies at order $S^2$. If we include quantum corrections to the ground state energy only at order $S$, the estimate for the classical spin flip shifts to the much smaller value $h^*/J_b\simeq 0.42$, indicated as dashed vertical line in Fig.~2(a) of the main text. Moreover, the latter estimate neglects a possible spin canting in the classical configuration near the flipped spin. (In the adatom case, this can only happen if the bulk spin is flipped.) We stress that, while this semiclassical analysis tells us that a spin flip  must occur as we decrease $h/J_K$, the comparison with ED indicates that the discontinuous spin transition of a $S_{\rm imp}=1/2$ magnetic impurity is better described by quantum fluctuations on top of the uniform spin configuration. 

\new{
\subsection{E. Difference to spin-flop transitions} \label{sec1e}
}

\new{We here describe the difference between the spin-flip transitions 
discovered in this work and the well-known spin-flop transition in an applied field \cite{Blundell}.  
The ``classical'' spin-flop mechanism occurs, e.g., for an antiferromagnetic XXZ model in an external field applied along the direction of the sublattice magnetization. For an easy-axis anisotropy,  
$J_z > J_{x,y}$, the low-field state remains a N{\'e}el state along the $z$-axis. 
At a critical field $h_c$, the spin flops to the canted state in which the spins now have 
a projection in the $x$-$y$ plane that spontaneously breaks the spin rotation symmetry along the
$z$-axis. This transition is of first order, and there is no reason for the magnon gap 
to close due to the anisotropies. In general, one expects a gap closing only at the 
transition to the polarized phase.}

\new{The situation described in our work is very distinct. We are in a (partially) polarized 
phase, where spin-flip transitions only take place in the vicinity of a magnetic quantum impurity. 
In the adatom position,  for instance, the impurity and the bulk spin that interact antiferromagnetically cannot acquire in-plane components nor rotate continuously toward the
polarized state because the local interactions in the quantum impurity problem cannot spontaneously break the (discrete) rotation symmetry along the $\hat{\mb c}$ direction. 
In contrast to the mean-field-like argument of the spin-flop transition \cite{Blundell}, the local spin-flip transitions described here involve the gap closing of a magnon bound state, 
which then leads to jumps in the local magnetization.}\\

\section{II. Analytical results for bound state energies}\label{sec2}

In this section, we consider an adatom impurity position and a perpendicular magnetic field,
$\mb h=\hat{\mb c}$.  We study the KH model in the weak-coupling limit 
$J_K\ll h$. \new{As shown below, we arrive at a simple structure of the resulting continuum theory 
describing the magnon bound state and the associated spin-flip transition whenever the bound state energy crosses zero.  This continuum argument shows that the predicted spin-flip transitions can occur in general 2D magnets, i.e., not only in van der Waals magnets described by the KH honeycomb model. Moreover, repeating similar arguments for 1D spin chains (and to some extent, even 
3D magnets), spin-flip transitions are expected to occur as well.
In our example, } there is no spin canting and we stay away from spin transitions. 
Our goal is to obtain analytical results for the sub-gap magnon bound state energy.

Under the above conditions, we can treat the adatom as a weak perturbation to the homogeneous system. We take the continuum limit of the LSW Hamiltonian \eqref{lsw} by expanding the free magnon spectrum about the band  minimum at the $\Gamma$ point. The  magnon dispersion is then approximated by $\omega(\mb k)\approx \omega_g+\frac{\mb k^2}{2m}$, where $\omega_g$  is the magnon gap and  the mass term is isotropic. \new{We here consider a partially polarized phase without spontaneous symmetry breaking. While anisotropic exchange interactions such as the Kitaev interaction $K$
may contribute to $\omega_g$ in general,
in the partially polarized phase of the KH model, 
the magnon gap increases with the magnetic field according to $\omega_g=h$ without 
contributions due to $K$. Moreover, we find} 
\be\label{meff}
    m^{-1}=J+\frac{K}{3}-\frac{K^2}{h+3J+K}.
\ee 
 We recall that for $K\ne 0$, magnon number is not conserved, see Eq.~\eqref{lsw}. In the unperturbed free-magnon Hamiltonian, this effect is taken care of by the Bogoliubov transformation that gives the magnon dispersion to order $k^2$. Next, we take the continuum limit of the impurity contribution, where we expand the bulk modes in terms of momentum eigenstates and neglect terms beyond zeroth order in $k$. As a result, the Kondo interaction and the   Zeeman term for the impurity spin give
\bea\nonumber
   H_{\rm imp} &=& -J_KS_{\rm{imp}} b^\dagger(\mb R_1) b(\mb R_1)\nonumber\\
   &&+J_K\sqrt{SS_{\rm{imp}}}[b^\dagger(\mb R_0) b(\mb R_1)+\text{h.c.}] \nonumber \\
   &&+(gh-SJ_K)b^\dagger(\mb R_0) b(\mb R_0)+\rm{const.},
\eea
where $b(\mb R)$ annihilates a boson at position $\mb R$, with $\mb R_0$ and $\mb R_1$ being the positions of the impurity and the bulk spin coupled to the impurity, respectively. We single out the impurity mode by introducing the notation $a=b(\mb R_0)$, whereas all the bulk modes are represented by $b(\mb r)$ with $\mb r\in\mathbb R^2$ a vector in the plane that contains the honeycomb lattice. Setting $\mb R_1=0$ and measuring the energies with respect to the lower threshold $\omega_g$ of the magnon continuum, we   write the  effective low-energy Hamiltonian in real space as
\bea \nonumber
H_{\rm eff}&=& \int d^2r\, b^\dagger(\mb r) \left[ -\frac{\nabla^2}{2m}-V_0\delta(\mb r)\right]b(\mb r) \\ 
&+&\varepsilon_a a^\dagger a +w\left[a^\dagger b(\mb r=0)+ b^\dagger (\mb r=0)a\right].\label{Heff}
\eea
Here, $V_0= J_KS_{\rm imp}$ is the strength of an attractive potential  for the bulk  magnons induced by the magnetization of the impurity. 
The energy of the bosonic state associated with the impurity  (measured from the bulk magnon gap $\omega_g$) is 
\be
\varepsilon_a=gh-\omega_g - J_K S.
\ee  
Note that the occupation $a^\dagger a$ of this state can change due to quantum fluctuations of the impurity.
Finally, $w= J_K\sqrt{S S_{\rm imp}}$ is an effective hybridization due to the transverse part of the Kondo interaction.
The classical magnetic impurity limit \cite{Imry1975,Bauer2023} corresponds to taking  $S_{\rm imp}\to\infty$ and $J_K\to0$ while keeping the product $V_0=J_K S_{\rm imp}$ constant. In this limit, one finds $w\to 0$, i.e., the $a$ boson decouples from the $b({\mb r})$ bosons. Then the only effect of the impurity is to act as a static local field on the bulk degrees of freedom.  

Noting that the effective Hamiltonian \eqref{Heff} conserves the total magnon   number, we use an \emph{Ansatz} for a general single-particle state,
\be
\left|\Psi\right\rangle=\left(\int d^2r\, \psi(\mb r)b^\dagger(\mb r)+\phi_a a^\dagger\right)|0\rangle,
\ee
where $\psi(\mb r)$ and $\phi_a$ are wave functions and $|0\rangle$ is the vacuum state, identified with the ground state of the unperturbed Hamiltonian (for $J_K=0)$. 
The Schr\"odinger equation, $H_{\rm eff}\left|\Psi\right\rangle=E\left|\Psi\right\rangle$, where the energy $E$ is measured relative to $\omega_g$, yields the coupled equations
\bea\nonumber
-\frac{1}{2m}\nabla^2\psi(\mb r) -V_0\delta(\mb r)\psi(\mb r)+w\phi_a\delta(\mb r)&=&E\psi(\mb r),\label{bs1}\\
\varepsilon_a\phi_a+w\psi(0)&=&E\phi_a.\label{bs2}
\eea
From Eq.~(\ref{bs2}), we observe that a bound state solution with $E=E_b<0$ satisfies the relation
\be\label{aux1}
\phi_a=-\frac{w\psi(0)}{\varepsilon_a+|E_b|}.
\ee
Substituting Eq.~\eqref{aux1} into the first equation in Eq.~(\ref{bs1}), we obtain the Schr\"odinger equation \new{(M3)} for a $\delta$-function potential with a renormalized coupling that depends on the energy itself,
\be\label{geff}
V_{\rm eff}(E_b)=V_0+\frac{w^2 }{\varepsilon_a+|E_b|}. 
\ee

For a \emph{classical} magnetic impurity, $w\to 0$ implies that $V_{\rm eff}= V_0$ is constant.
Hence \new{Eq.~(M3)} can be solved after imposing a large-momentum cutoff $\Lambda$ \cite{Jackiw1991}. The resulting bound state energy is exponentially small at weak coupling but depends on $V_0$ and $m$ in a nonperturbative manner, 
\be\label{classicalbs}
E_b \simeq -\frac{\Lambda^2}{2m}e^{-2\pi/(m V_0)}.
\ee
 
\begin{figure}[t]
\begin{center}
\includegraphics[width=\columnwidth]{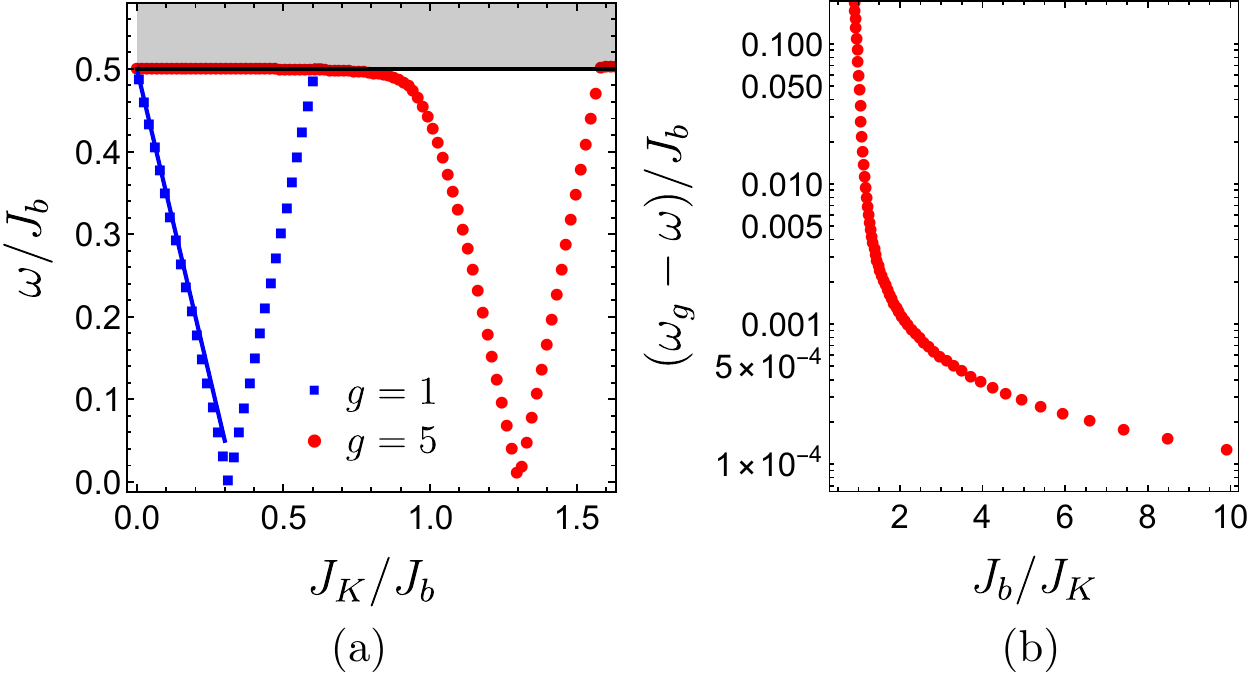} 
\end{center}
\caption{Magnon spectrum $\omega$ for the KH model with an adatom quantum magnetic impurity 
vs impurity coupling $J_K$ (both $\omega$ and $J_K$ are in units of the bulk exchange coupling $J_b>0$). 
We set $S=3/2$, $S_{\rm imp}=1/2$, $h/J_b=0.5$, and $K/J_b=-0.2$, where the field is along the $\hat{\mb c}$ direction.
(a) The symbols represent the bound state energy calculated by LSW theory for $g=1$ (blue squares) and $g=5$ (red dots). The solid line for $g=1$ gives the analytical prediction (\ref{lineardep}). The shaded region represents the magnon continuum $\omega>\omega_g=h$. (b) Binding energy for $g=5$ in the weak-coupling regime, plotted on a logarithmic scale vs $J_b/J_K$. 
}
\label{figSM2}
\end{figure}

For a \emph{quantum impurity}, on the other hand, we obtain a qualitatively different scaling of the bound state energy. In this case,  $V_{\rm eff}(E_b)$ becomes large if $\varepsilon_a$ is negative and  
$\varepsilon_a\approx E_b$. In particular, for the KH model with  $g= 1$, we have $\omega_g=h$ and then $\varepsilon_a=-J_KS<0$ for any $J_K>0$.  In general, the binding energy is given by the solution of the equation 
\be
\frac{2m|E_b|}{\Lambda^2}=\exp\left[-\frac{2\pi}{m V_{\rm eff }(E_b)}\right],\label{generalEb}
\ee
In the regime $|V_{\rm eff }(E_b)|\gg 1$, we can approximate
\be
\frac{2m|E_b|}{\Lambda^2}\approx 1-\frac{2\pi}{mV_0+\frac{mw^2 }{\varepsilon_a+|E_b|}}, 
\ee
which is a quadratic equation in $|E_b|$.  The low-energy solution is 
\be
E_b=\varepsilon_a+\mc O(w^2)=   gh-\omega_g-J_K S+\mc O(J_K^2), \label{lineardep}
\ee
\new{as quoted in the main text.}
For $g=1$, the binding energy vanishes for $J_K\to0$ and scales linearly with $J_K$ at weak coupling. 
The linear dependence predicted by Eq.~(\ref{lineardep}), without any free fitting parameter,
is in excellent agreement with our numerical results for the lattice model, see Fig.~\ref{figSM2}(a). 
For $g>1$, the binding energy is exponentially small at weak coupling, see Fig.~\ref{figSM2}(b), but it
crosses over to the linear dependence as one increases $J_K$. The condition  $w^2/\varepsilon_a\sim V_0$, 
see Eq.~(\ref{geff}),  gives the estimate $J_K\sim \frac{(g-1)h}{2S}$ for the crossover scale. In the strong-coupling regime, i.e., above the spin transition, the bound state eventually merges with the continuum and disappears for sufficiently large $J_K$. 

 \section{III. On the STS conductance}\label{sec3}

In this section, \new{we summarize known results for the STS conductance  which have been used for generating Fig.~4 in the main text. For the tip position $\mb r$ and bias voltage $V$, the zero-temperature STS conductance can be written as \cite{fransson2010theory,fernandez2009theory,Feldmeier2020,Carrega2020,Mitra2023,Bauer2023,Kao2024a,Kao2024b}
\be\label{STScond}
    G(V) = G_0\sum_{j,k}\frac{t_jt_k^*}{|t_{\rm bulk}|^2}F_{j,k}(\mb r)  \sum_\alpha\int_0^{eV} d\omega\, C^{\alpha,\alpha}_{j,k}(\omega),
\ee
where $t_{j=0}=t_{\rm{imp}}$ and $t_{j\neq 0}=t_{\rm{bulk}}$ are cotunneling amplitudes discussed below. 
This expression holds in the weak-coupling limit, where  the tip (and the substrate) are kept sufficiently distant from both the impurity and the bulk spins in order to allow for a perturbative treatment of the respective tunneling processes.
The conductance reference scale used in Eq.~\eqref{STScond} is defined as 
\be \label{G0ref}
G_0=2\pi d_Ad_B |t_{\rm bulk}|^2,
\ee
where $d_A$ ($d_B$) is the density of states of the tip (substrate) at the Fermi level. 
Moreover, the function 
\be\label{Fjk} 
F_{j,k}(\mb r)=\exp\left(-\frac{|\mb{R}_j-\mb{r}|+|\mb{R}_k-\mb{r}|}{l_0}\right)
\ee
encodes the tip position dependence.
The spin sites (including the impurity) are denoted by ${\mb R}_j$, and $l_0$ is a length characterizing the STS resolution. 
Finally, the dynamical spin correlation function of the magnet (in the absence of tip and substrate) is given by 
\be\label{spinspincor}
C^{\alpha,\beta}_{j,k}(\omega)=\int dt\,{e}^{i\omega t}\langle \Phi_0|S^\alpha_j(t)S^\beta_k(0)|\Phi_0\rangle,
\ee
using the ground state $|\Phi_0\rangle$ of the 2D magnet.  
We note that on top of the inelastic contribution \eqref{STScond}, there
is a featureless voltage-independent elastic cotunneling term \cite{Feldmeier2020,Bauer2023}.
For finite bound state energy, the zero-bias conductance obtained from Eq.~\eqref{STScond} is expressed in terms of spin expectation values $\langle S^\alpha_j\rangle$ only,
\be\label{zerocond}
G(V=0)=G_0\sum_{\alpha}\biggl|\sum_{j}\frac{t_j}{|t_{\rm bulk}|}
e^{-|\mb{R}_j-\mb{r}|/l_0}\langle S^\alpha_j\rangle\biggr|^2.
\ee
We conclude that local spin-flip transitions are directly visible as steps in the zero-bias STS conductance.}

\new{
Let us briefly describe the main ideas behind the derivation of Eq.~\eqref{STScond}.}
In order to describe electrical transport
through a 2D magnet, one has to start from a model that retains the charge degrees of freedom, e.g., the Hubbard-Kanamori model.
By a Schrieffer-Wolff transformation to the low-energy spin sector, one then obtains the Hamiltonian (M1) in terms of the spin operators
${\mb S}_j$.  We now include tunneling processes with amplitude $t_A({\mb r}-{\mb R}_j)$ connecting a normal-conducting scanning probe tip at position ${\mb r}$, described by
noninteracting fermion operators $c_{A,\tau}({\mb r })$ for spin projection $\tau\in \{\uparrow,\downarrow\}$, to lattice site ${\mb R}_j$ of the 2D layer. 
Here we assume that $t_A$ is independent of the momentum and the spin of the tunneling electrons.
Similarly, we include tunneling processes with constant  amplitude $t_B$ connecting the respective site in the 2D layer to a 2D substrate described by noninteracting fermion operators $c_{B,\tau}({\mb R}_j)$. Below, Pauli matrices ${\boldsymbol\tau}=(\tau_x,\tau_y,\tau_z)$ and the
identity $\tau_0$ act in conduction electron spin space.
Applying a Schrieffer-Wolff transformation in the presence of these tunneling terms now
yields an effective cotunneling Hamiltonian connecting the tip and the substrate via the 2D magnetic layer, see \cite{fransson2010theory,fernandez2009theory,Feldmeier2020,Mitra2023,Bauer2023} for details,  
\bea \nonumber
H_{\rm cot} &=& \sum_j \frac{t^*_A({\mb r}-{\mb R}_j) t_B}{U} c_{A}^\dagger({\mb r}) \bigl( \eta_0 \tau_0 \mathbbm{1}_j +
\\ &+& \eta_1 {\boldsymbol \tau}\cdot {\mb S}_j \bigr) c_B^{}({\mb R}_j) + {\rm H.c.}, \label{cot}
\eea
where the sum runs over all lattice sites of the 2D magnet in Eq.~(M1), including the impurity site. Here, $U$ is a Coulomb interaction energy
scale of the 2D layer, and $c_{A/B}^\dagger=(c_{A/B,\uparrow}^\dagger,c_{A/B,\downarrow}^\dagger)$.
The numbers $\eta_0$ and $\eta_1$ depend, in particular, on the ratio of the Hund's rule 
coupling $J_H$ and the parameter $U$, and are of order unity \cite{Bauer2023}. (In general, they are different for bulk and
impurity sites.) We note in passing that in general there  is also 
a spin-rotation-symmetry breaking contribution \cite{Bauer2023} which we have neglected in Eq.~\eqref{cot}. 
We assume an exponential distance dependence of the tunneling matrix elements, $t_A({\mb r}-{\mb R}_j)\propto e^{-|{\mb r}-{\mb R}_j|/l_0}$, where the length scale $l_0$ sets the resolution  of the scanning probe tip.  
We also assume a constant and spin-independent (non-polarized) density of states in the relevant energy range for both the tip ($d_A$) and the substrate ($d_B$). 

A computation of the current $I(V)$ from tip to substrate, and thus of the differential conductance $G(V)=dI/dV$, can now be performed by perturbation theory in $H_{\rm cot}$. To that end, we proceed along standard steps, see, e.g., Ref.~\cite{Bauer2023}. 
Within Fermi's golden rule, the squared matrix elements of $\eta_0 \tau_0 \mathbbm{1}_j +
\eta_1 {\boldsymbol \tau}\cdot {\mb S}_j$ appear. The term $G\propto \eta_0^2$ is associated with elastic tunneling processes and gives rise to a background conductance which is independent of voltage and tip position. 
The term $G\propto\eta_0\eta_1$ is also independent of voltage but generally depends on the tip position as it involves the spin expectation values $\braket{S^\alpha_j}$.
However, it vanishes under the above assumption of a non-polarized tip and substrate.
Finally, the term $G\propto \eta^2_1$ depends on voltage in a nontrivial manner since it probes the dynamical spin correlations of the 
2D magnet. This term is shown in \new{Eq.~\eqref{STScond}}.
Accounting for the exponential distance dependence of $t_A$ through the form 
factors $F_{j,k}$ in \new{Eq.~\eqref{Fjk}}, all other microscopic factors in Eq.~\eqref{cot} 
appear only via the overall scale $G_0$ \new{in Eq.~\eqref{G0ref}} for the cotunneling conductance, and through  cotunneling amplitudes $t_j$.  For the latter, we distinguish between bulk sites (with $t_{j\neq 0}=t_{\rm bulk}$) and the 
impurity site (with $t_{j=0}=t_{\rm imp}$).  The ratio $t_{\rm imp}/t_{\rm bulk}$ may
vary substantially  depending on the impurity type (adatom vs substitutional). In  Fig.~3 of the main text, we considered $t_{\rm imp}>t_{\rm bulk}$, assuming a stronger overlap of the tip wavefunction with Co adatoms than with Cr$^{3+}$ ions in the honeycomb layer.  After the above steps,  one arrives at \new{Eq.~\eqref{STScond}}.

The conductance is thereby expressed in terms of dynamical spin correlations of the 2D magnet in the absence of tip and substrate, see Sec. I\,C.  These correlations exhibit a peak at the bound state energy which translates to a step in the conductance.
As discussed in the main text, one can thereby track the energy of the bound state as a function of the external magnetic field by measuring $G(V)$ or $dG/dV$.   
Moreover, $G(V=0)$ can directly monitor atomic-scale spin-flip transitions. 

\section{IV. Kitaev honeycomb model}\label{sec4}

\begin{figure}
    \centering
    \includegraphics[width=0.98\columnwidth]{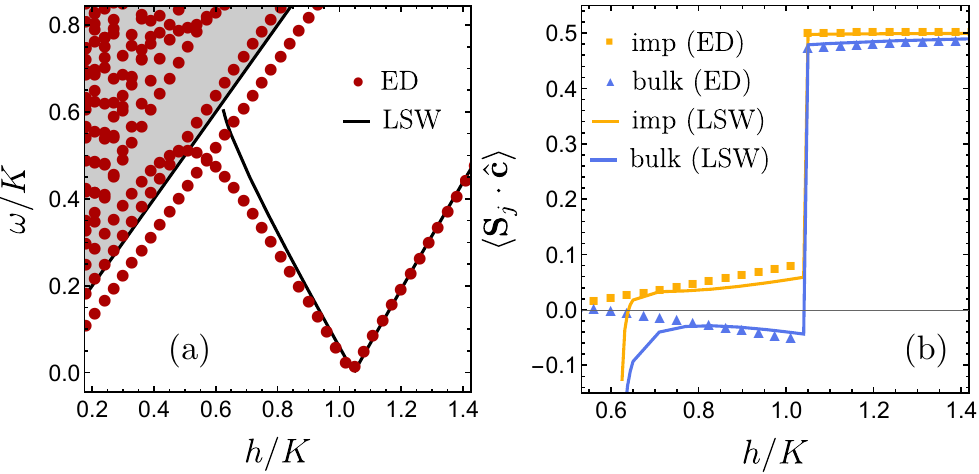}
    \caption{ED and LSW results for a magnetic  quantum impurity in the adatom configuration for the ferromagnetic Kitaev model ($K>0$, $J_b=0$) with $S=S_{\rm imp}=1/2$, 
$J_K/K=1.5$, $g=1.5$, and  $\mb h=h\hat{\mb c}$. (a) Excitation spectrum vs magnetic field.  ED (LSW) results correspond to red dots (black lines). ED results were obtained 
for a system of $2\times 2$ unit cells with periodic boundary conditions. (b) Spin projections $\langle{\mb S}_{0,1}\cdot \hat{\mb c}\rangle$ vs $h/J_b$.
Yellow squares (lines) correspond to ED (LSW) results for the impurity spin. Results for the
coupled bulk spin are shown as blue triangles (ED) and blue lines (LSW), respectively. 
}
    \label{figSM3}
\end{figure}

We here show results for the ferromagnetic 2D Kitaev honeycomb model with $J_b=0$ and $K>0$ \cite{Kitaev2006} in the partially polarized phase, which may serve as idealized model for $\alpha$-RuCl$_3$ in a strong magnetic field. For the bulk spins (Ru$^{3+}$), we effectively have $S=1/2$. 
Using Co atoms as examples for the adatom magnetic impurities \cite{Chen2022}, we have $S_{\rm imp}=1/2$ with $g\approx 1.5$. 
 In contrast to Fig.~2(b) of the main text, we now find that the discontinuity in the local magnetization at the transition ($h/K\approx 1.05$) is larger for the bulk than for the impurity spin. In fact, the magnetization of the bulk spin becomes slightly negative for magnetic fields just below the transition. While the spin flip of the bulk spin is
 expected to occur for the large-$g$ limit,  we here observe that it can happen already for moderate values of $g>1$.  Despite of the small system size used in the ED calculations, 
 LSW and ED results agree rather well for large magnetic fields.  However, Fig.~\ref{figSM3} shows that they significantly deviate for $h/K<0.7$, i.e., once the bound state approaches the continuum. However, since the main emphasis of our  work is on spin-flip transitions associated with zero-energy bound states, we do not study this regime here in detail. 
 
 \begin{figure}
    \centering
    \includegraphics[width=0.98\columnwidth]{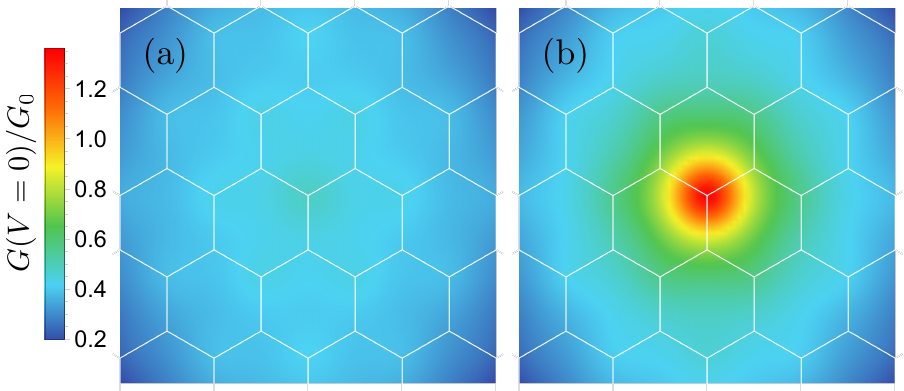}
    \caption{Spatial profile of the zero-bias STS conductance near an adatom magnetic impurity for the ferromagnetic Kitaev model, with the same parameters as in Fig.~\ref{figSM3}.
    Results obtained from LSW theory  for two field values: (a) $h = 0.854K$ (below the transition), (b) $h = 1.33K$ (above the transition).}
    \label{figSM4}
\end{figure}

In Fig.~\ref{figSM4}, we show the spatial profile of the STS zero-bias conductance for the Kitaev model near the adatom impurity studied in Fig.~\ref{figSM3}. 
Similar to our results for the KH model in Fig.~3 of the main text, the zero-bias conductance is significantly enhanced in the vicinity of the impurity 
for magnetic fields above the transition. In the present case, however, both the impurity and the coupled bulk spin show positive magnetization. In fact,
below the critical field, the conductance has a local maximum when the tip is right above the impurity, in contrast to the local minimum observed 
in Fig.~3(c) of the main text.  This difference can be rationalized by noticing that the parameter choice $t_{\rm imp}>t_{\rm bulk}$ used in Fig.~\ref{figSM4} favors 
contributions due to cotunneling via the impurity site.  The corresponding impurity spin still has a small positive magnetization even below the transition, see Fig.~\ref{figSM3}(b),
in contrast to the negative magnetization observed below the transition in Fig.~2(b) of the main text.

\end{document}